\documentclass[conference, letterpaper]{IEEEtran}
\IEEEoverridecommandlockouts

\usepackage[noadjust]{cite}
\usepackage{amsmath,amssymb,amsfonts}
\usepackage{algorithmic}
\usepackage{graphicx}
\usepackage{textcomp}
\usepackage{xcolor}
\usepackage{siunitx}
\usepackage[nolist,nohyperlinks]{acronym}

\usepackage{float}
\usepackage{stfloats}
\usepackage{tabularx}
\usepackage{mathtools}
\usepackage{booktabs}
\usepackage{multirow}
\usepackage{svg}
\usepackage[percent, Option]{overpic}
\usepackage{tikz}
\usepackage{placeins}
\usepackage{orcidlink}
\usepackage[capitalise]{cleveref}
\usepackage{siunitx}
\usepackage[acronym]{glossaries}
\usepackage{utfsym}
\usepackage{threeparttable}
\usepackage{caption}
\usepackage[normalem]{ulem}

\usetikzlibrary{shapes,arrows}
\usetikzlibrary{circuits.logic.US} 

\usepackage{fancyhdr}

\usepackage[utf8]{inputenc}

\usepackage[acronym]{glossaries}

\usepackage[inline]{enumitem}

\newacronym{UWB}{UWB}{ultra-wideband}
\newacronym{MEMS}{MEMS}{microelectromechanical systems}
\newacronym{IoT}{IoT}{Internet of Things}
\newacronym{NLoS}{NLOS}{non-line-of-sight}
\newacronym{LoS}{LOS}{line-of-sight}
\newacronym{IMU}{IMU}{inertial measurement unit}
\newacronym{GNSS}{GNSS}{global navigation satellite system}
\newacronym{WSN}{WSN}{wireless sensor network}
\newacronym{ToF}{ToF}{time-of-flight}
\newacronym{RSSI}{RSSI}{received signal strength indicator}
\newacronym{RF}{RF}{radio frequency}
\newacronym{MCU}{MCU}{microcontroller}
\newacronym{IC}{IC}{integrated circuit}
\newacronym{LRP}{LRP}{low-rate pulse repetition}
\newacronym{HRP}{HRP}{high-rate pulse repetition}
\newacronym{PHY}{PHY}{physical interface}
\newacronym{EVK}{EVK}{evaluation kit}
\newacronym{OOK}{OOK}{on-off keying}
\newacronym{2bit-PPM}{2bit-PPM}{2bit-pulse-position modulation}
\newacronym{PDR}{PDR}{packet delivery ratio}
\newacronym{BLE}{BLE}{Bluetooth Low Energy}
\newacronym{GUI}{GUI}{graphical user interface}
\newacronym{TWR}{TWR}{two-ways ranging}
\newacronym{AoA}{AoA}{angle of arrival}
\newacronym{PDoA}{PDoA}{phase difference of arrival}
\newacronym{MCPD}{MCPD}{multi-carrier phase-difference}
\newacronym{SoC}{SoC}{system-on-chip}
\newacronym{PCB}{PCB}{printed circuit board}

\usepackage{eso-pic}
\usepackage{url}
\AddToShipoutPictureBG*{
  \AtPageUpperLeft{%
    \put(0,-40){\raisebox{15pt}{\makebox[\paperwidth]{\begin{minipage}{21cm}\centering
      \textcolor{gray}{This article has been accepted for publication in the proceedings of the 2023 IEEE SENSORS conference.\\
      DOI: 10.1109/SENSORS56945.2023.10325219} 
    \end{minipage}}}}%
  }
  \AtPageLowerLeft{%
    \raisebox{25pt}{\makebox[\paperwidth]{\begin{minipage}{21cm}\centering
      \textcolor{gray}{ \copyright 2023 IEEE.  Personal use of this material is permitted.  Permission from IEEE must be obtained for all other uses, in any current or future media, including reprinting/republishing this material for advertising or promotional purposes, creating new collective works, for resale or redistribution to servers or lists, or reuse of any copyrighted component of this work in other works.
      }
    \end{minipage}}}%
  }
}

\begin{document}
\bstctlcite{IEEEexample:BSTcontrol} 

\title{Evaluation of a Non-Coherent Ultra-Wideband Transceiver for Micropower Sensor Nodes}

\definecolor{darkred}{HTML}{A8322C}
\newcommand{\revA}[1]{\textcolor{black}{#1}}

\author{
\IEEEauthorblockN{
Jonah Imfeld~\orcidlink{0009-0009-1413-9824}, 
Silvano Cortesi~\orcidlink{0000-0002-2642-0797},
Philipp Mayer~\orcidlink{0000-0002-4554-7937},
Michele Magno~\orcidlink{0000-0003-0368-8923}}
\IEEEauthorblockA{\textit{Dept. of Information Technology and Electrical Engineering, ETH Z\"{u}rich, Zürich, Switzerland}}
}


\markboth{Journal of \LaTeX\ Class Files,~Vol.~14, No.~8, August~2015}%
{Shell \MakeLowercase{\textit{et al.}}: Bare Demo of IEEEtran.cls for IEEE Journals}

\maketitle

\begin{abstract}\label{sec:abstract}
Spatial and contextual awareness has the potential to revolutionize sensor nodes, enabling spatially augmented data collection and location-based services. 
With its high bandwidth, superior energy efficiency, and precise time-of-flight measurements, \gls{UWB} technology emerges as an ideal solution for such devices.

This paper presents an evaluation and comparison of a non-coherent UWB transceiver within the context of highly energy-constrained wireless sensing nodes and pervasive \gls{IoT} devices. Experimental results highlight the unique properties of \gls{UWB} transceivers, showcasing efficient data transfer ranging from \(\SI{2}{\kilo\bit/\s}\) to \(\SI{7.2}{\mega\bit/\s}\) while reaching an energy consumption of \(\SI{0.29}{\nano\joule/\bit}\) and \(\SI{1.39}{\nano\joule/\bit}\) for transmitting and receiving, respectively. Notably, a ranging accuracy of up to \(\pm\SI{25}{\centi\meter}\) can be achieved. Moreover, the peak power consumption of the \gls{UWB} transceiver is with \(\qty{6.7}{\milli\watt}\) in TX and \(\qty{23}{\milli\watt}\) in RX significantly lower than that of other commercial \gls{UWB} transceivers.

\end{abstract}
\vspace{10pt}
\begin{IEEEkeywords}
Wireless communication, ultra-wideband (UWB), Internet of Things (IoT), localization, ultra-low power, sensor systems and applications.
\end{IEEEkeywords}
\glsresetall

\section{Introduction}\label{sec:introduction}
The increasing adoption of \gls{IoT} sensors and devices has sparked a demand for pervasive location services. In addition to prominent applications such as navigation and asset tracking, the inclusion of location awareness in sensors has the intriguing ability to expand the horizons of mobile IoT systems\cite{Zou2016, Li2021}. It empowers them to interact with the surrounding environment and potentially paves the way for contextual understanding, opening new possibilities for innovation and sensor interaction \cite{Elsharkawy21, mayer2022selfsustaining, Zhao2023}.

While accurate outdoor localization is achievable using the \gls{GNSS}, indoor and hybrid scenarios lack widely adopted solutions. Indoor location services present a larger market potential, thus driving research efforts toward \gls{GNSS}-like indoor localization technologies \cite{Cheema18}.

\Gls{UWB} based on IEEE 802.15.4z has emerged as a leading choice for high-accuracy device-based indoor localization~\cite{ibnatta21_expos_evaluat_differ_indoor_local_system}. With its sub-nanosecond pulses and the resulting superior time resolution, \gls{UWB} enables precise \gls{ToF} measurements \cite{flueratoru22_high_accur_rangin_local_with}, allowing for accurate distance estimation and ranging. Alongside the ranging capabilities, \gls{UWB} has the potential for highly energy-efficient communication, particularly on the transmitter side and when utilizing non-coherent receiver implementations, making it a promising technology for battery- or energy-harvesting powered devices \cite{Dardari20, Bauwens2021, Jeon21}.  

One significant drawback of commercial \gls{UWB} transceivers is their high peak power consumption, which presents challenges for small-sized battery-powered devices. In addition, it is crucial to eliminate the need for periodic synchronization between these devices and infrastructure and to achieve near zero watts idle consumption to enable long-term operation without frequent recharging or replacement \cite{mayer2022selfsustaining}. To incorporate the ability of accurate ranging and efficient communication in a single device, a hybrid wireless frontend is often implemented that uses both \gls{BLE} and \gls{UWB}~\cite{kolakowski20_uwb_ble_tracking_system, coppens22_overv_uwb_stand_organ_ieee}. 

This work comprehensively and accurately evaluates the \textsc{Spark Microsystems SR1000} series, a promising ultra-wideband transceiver that combines key features suitable for highly energy-constrained systems. In particular, the main contributions of this work are as follows:
\begin{enumerate*}[label=(\roman*),font=\itshape]
\item A comprehensive analysis of the energy consumption and the optimal configuration to balance data rate and power consumption;
\item An accurate survey on recently released commercial \gls{UWB} transceivers;
\item An evaluation of the ranging capability, demonstrating a ranging accuracy of up to \(\pm\SI{25}{\centi\meter}\). 
\end{enumerate*}

The rest of the paper is organized as follows: 
Section \ref{sec:relatedWork} presents the recent literature and discusses key performance indicators of commercial \gls{UWB} transceivers;
Section \ref{sec:setup} introduces the experimental setup;
Section \ref{sec:results} shows experimental results and possible applications;
Finally, section \ref{sec:conclusion} concludes the paper.

\section{Related Work}\label{sec:relatedWork}
In addition to \gls{UWB}, \gls{BLE} stands out as one of the most widely adopted technologies for low-power data transmission and positioning. With the standardization of \gls{BLE} direction finding and its availability in modern and low-cost \glspl{SoC} such as the nRF5340, \gls{BLE} showed itself effective for localization through \gls{AoA} and \gls{MCPD} \cite{cortesi23_compar_between_rssi_mcpd_based}. Depending on the filters used, the ranging accuracy is below \SI{60}{\centi\meter}~\cite{mackey20_improv_ble_beacon_proxim_estim, zhu18_rf_techn_indoor_local_posit}, while an \gls{AoA} accuracy of \ang{10} can be achieved at \SI{5}{\meter}~\cite{langberg20_aoa}. The power consumption of \gls{BLE} makes it particularly suited for micropower sensors, demonstrating  \SI{11.2}{\milli\watt} and \SI{9.5}{\milli\watt} for TX and RX, respectively, with a data rate of \SI{2}{\mega\bit/\second}~\cite{nrf5340ds}. An overview of the state-of-the-art commercial \gls{UWB} transceivers and their comparison against \gls{BLE} is given in \cref{tab:comparison}. 

In~\cite{coppens22_overv_uwb_stand_organ_ieee}, Coppens et al. give an overview of the current state of \gls{UWB} modules and standards. The IEEE 802.15.4z standard is supported by all popular \glspl{IC} except the SR1000, as shown in \cref{tab:comparison}. The devices based on IEEE 802.15.4z can be split into those supporting \gls{HRP} (\textsc{Qorvo} DW3xxx, \textsc{NXP} SR040 \& SR150) and those supporting \gls{LRP} (\textsc{Microchip} ATA8352). As shown in~\cite{margiani23_angle_arriv_centim_distan_estim, polonelli22_perfor_compar_decaw_dw100_dw300}, the DW3x20 devices achieve an accuracy of \ang{5} for \gls{AoA} and below \SI{5}{\centi\meter} at a power consumption of \SI{55}{\milli\watt}. \cite{juran22_hands_exper_uwb, heinrich23_smart_with_uwb} characterized the \textsc{NXP} SR040 and SR150 and found its accuracy to be below \ang{8} or \SI{15}{\centi\meter}. An evaluation of Flueratoru et al.~\cite{flueratoru22_high_accur_rangin_local_with} showed that the \textsc{3db Access} 3DB6830C, the predecessor of the \textsc{Microchip} ATA8352, achieved a similar ranging accuracy of up to \SI{5}{\centi\meter} in \gls{LoS} conditions. The power consumption of the 3DB6830C is \SI{20.7}{\milli\watt} during TX and \SI{40.7}{\milli\watt} during RX. The increased energy efficiency of \gls{LRP}-based devices comes at the cost of a limited data rate of max. \SI{10}{\mega\bit/\second}~\cite{coppens22_overv_uwb_stand_organ_ieee}.
The SR1000 places itself in between \gls{LRP} and \gls{HRP}, with the purpose of attaining both low power consumption and high throughput.

\begin{table*}
\caption{Comparison of common \gls{UWB} and \gls{BLE} radio system-on-chips.}\label{tab:comparison}
\centering
\resizebox{\textwidth}{!}{%
\renewcommand{\arraystretch}{1.35}
\begin{tabular}{l|ccccc} 
\toprule
                                         & nRF5x BLE~\cite{mackey20_improv_ble_beacon_proxim_estim, zhu18_rf_techn_indoor_local_posit, langberg20_aoa, nrf5340ds}    & Qorvo DW3220~\cite{polonelli22_perfor_compar_decaw_dw100_dw300, margiani23_angle_arriv_centim_distan_estim, qorvo_ds}      & NXP SR040, SR150~\cite{juran22_hands_exper_uwb, heinrich23_smart_with_uwb, sr150_ds}       & Microchip ATA8352~\cite{flueratoru22_high_accur_rangin_local_with, microchip_ds, microchip_docs}        & Spark SR1000~\cite{sr1010_ds, sr1020_ds}   \\
\midrule
Frequency range [\SI{}{\GHz}]            & 2.4      & 6-8.5             & 6.2-8.2          & 6.2-8.3                     & \textbf{SR1010:} 3.1-5.8, \textbf{SR1020:} 6-9.3     \\ 
3\SI{}{\decibel} Bandwidth  [\SI{}{\GHz}]   & -         & 0.5               & 0.5              & 0.5                         & 3                                   \\
Max. Data rate  [\SI{}{Mbps}]            & 2            & 6.8               & 31.2                & 1                           & 10                                  \\
\midrule
Ranging / AoA Type                       & RSSI, MCPD, I/Q         & TWR, TDoA, PDoA   & TWR, TDoA, (PDoA)        & TWR, TDoA                   & TWR                                 \\
Ranging Accuracy [\SI{}{\centi\meter}]   & \(\pm 60\)            & \(\pm 5\)             & \(\pm 10\)           & \(\pm 5\)                       & \(\pm 30\)                               \\
AoA Accuracy [\SI{}{\degree}]            & \(\pm 10\)            & \(\pm 5\)            & \(\pm 3\)            & -                           & -                                   \\
\midrule
Shutdown Power draw [\SI{}{\nano\watt}]     & 1620         & 468                & 900                & 87                          & 181                                  \\
Peak TX Power draw [\SI{}{\milli\watt}]          & 11.2         & 63                & 236                & 30                          & \textbf{SR1010:} 6.7 , \textbf{SR1020:} 6.3   \\
Peak RX Power draw [\SI{}{\milli\watt}]          & 9.5          & 158                & 207                & 130                         & \textbf{SR1010:} 23 , \textbf{SR1020:} 21.2     \\
\midrule
IEEE802.15.4z Compliance                 & -            & \usym{2713}       & \usym{2713}      & \usym{2713}\tnote{1}       & \usym{2717}                         \\
UWB PHY Type                             & -            & HRP               & HRP              & LRP\tnote{1}               & Custom                              \\
Supported Channels                       & -            & 5 \& 9            & 5, 6, 8 \& 9     & -\tnote{1}                 & -                                   \\
FiRa Cerification                        & -            & \usym{2713}       & \usym{2713}      & \usym{2717}\tnote{1}       & \usym{2717}                         \\
\bottomrule
\end{tabular}
}
\vspace*{-1.25em}
\end{table*}
\section{Measurement Setup}\label{sec:setup}
The SR1000 \gls{EVK} from  \textsc{Spark Microsystems} is used as a platform to carry out all the measurements. It includes an STM32G473 \gls{MCU}, a lipo battery with a charging circuit, and interchangeable transceiver modules. The transceiver modules include the SR1010 or SR1020 transceiver and a monopole \gls{PCB} microstrip antenna. The specific modules used are the RMV8QMREV150 for the SR1010 and the RMV8QEUMMREV120 for the SR1020. Power measurements were conducted using a \textsc{Keysight} N6781A power analyzer connected to the module's current measurement header that is directly connected to the transceiver IC. Along with the \gls{EVK}, an application binary controllable by a \gls{GUI} is provided by \textsc{Spark}.

\section{Experimental Results}\label{sec:results}
\subsection{Power Consumption with Experimental Measurements}\label{sub:power}
Firstly, the static power consumption is analyzed. The transceivers allow five static configurations: shutdown mode, deep and shallow sleep, idle, or receiving. For each state, the current was measured at \SI{1.8}{\volt} and \SI{3.3}{\volt} by substituting the board supply with a source measurement unit. Deeper power states result in lower idle consumption but at the cost of higher wake-up time and energy.

The results of the static power consumption measurements are summarized in \cref{table:static_current_cons}. It should be noted that the startup cost from the shutdown state is around \SI{500}{\micro\joule} at \SI{1.8}{\volt} due to the need to recalibrate the transceiver for \SI{20}{\ms}. Consequently, entering the shutdown state becomes advantageous only if no message exchanges occur for a minimum period of \SI{644}{\s} for the SR1010 \gls{IC} at \SI{1.8}{\volt}.

\begin{table}[!b]
\vspace*{-1em}
\caption{Static current consumption as a function of the power mode and supply voltage.}
\centering
\renewcommand{\arraystretch}{1.35}
\begin{tabular}{@{}l*{4}{c}c@{}}
\toprule
\multirow{2}{*}{Transceiver} & \multicolumn{2}{c}{SR1010} & \multicolumn{2}{c}{SR1020}  \\
    & \SI{1.8}{\volt} & \SI{3.3}{\volt} &  \SI{1.8}{\volt} &\SI{3.3}{\volt}  \\
\midrule
Shutdown            & \SI{44}{\nano\ampere}         & \SI{100}{\nano\ampere}        & \SI{39}{\nano\ampere}         & \SI{104}{\nano\ampere}  \\ 
Deep Sleep          & \SI{475}{\nano\ampere}        & \SI{819}{\nano\ampere}        & \SI{466}{\nano\ampere}        & \SI{844}{\nano\ampere}  \\ 
Shallow Sleep       & \SI{39.1}{\micro\ampere}      & \SI{40.0}{\micro\ampere}      & \SI{39.5}{\micro\ampere}      & \SI{40.5}{\micro\ampere}  \\
Idle                & \SI{351.9}{\micro\ampere}     & \SI{263.6}{\micro\ampere}     & \SI{380.6}{\micro\ampere}     & \SI{269.2}{\micro\ampere}  \\
Active RX           & \SI{11.5}{\milli\ampere}     & \SI{7.2}{\milli\ampere}      & \SI{12.7}{\milli\ampere}     & \SI{7.7}{\milli\ampere}  \\
\bottomrule
\end{tabular}
\label{table:static_current_cons}
\end{table}
Secondly, the transmission and reception power consumption is measured in a data streaming configuration at \SI{1.8}{\volt}. The devices are set to transmit random data with \gls{OOK} modulation and deactivated forward error correction. Channel switching is used to maximize the link budget. The measurements are taken at an output power of \SI{0}{\dB FS} and \SI{-9}{\dB FS}. The data rate has been adjusted by varying the packet rate while maintaining a fixed payload size of \SI{124}{bytes}. As the minimum packet rate configurable through the \gls{GUI} is \SI{200}{packets/\s}, the payload size below has been reduced to accommodate data rates lower than \SI{0.198}{\mega bps}.

\begin{figure*}[!t]
    \centering
    \begin{overpic}[width=0.32\linewidth]{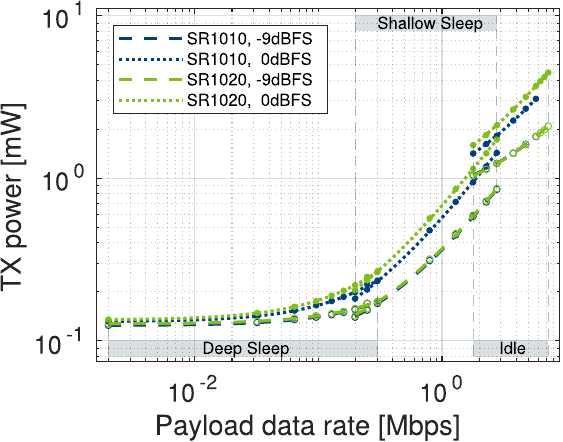}
        \put(5,1){(a)}
    \end{overpic}
    \begin{overpic}[width=0.32\linewidth]{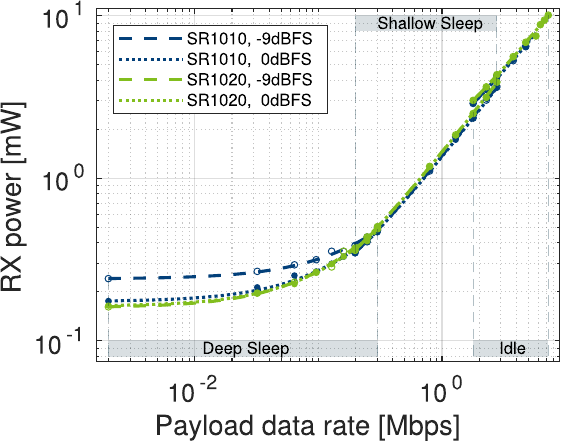}
        \put(5,1){(b)}
    \end{overpic}
    \begin{overpic}[width=0.32\linewidth]{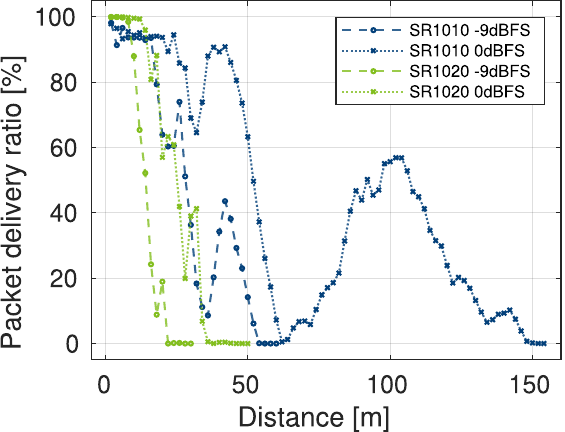}
        \put(5,1){(c)}
    \end{overpic}
        \caption{Power consumption during (a) TX and (b) RX data streaming. (c) Packet delivery ratio of the SR1000 \glspl{SoC}.}
        \label{fig:power_res}
        \vspace*{-1em}
\end{figure*}

The power consumption for transmission and reception in a datastream configuration is depicted in \cref{fig:power_res}a) and \cref{fig:power_res}b), respectively. The underlying power state is indicated within grey boxes. The relatively higher power consumption for SR1010 RX at \SI{-9}{\dB FS} in the deep sleep state is linked to a low packet delivery ratio which causes the receiver to remain active for longer.

Due to the high start-up energy cost of the deep sleep state, it is found to be less efficient compared to the shallow sleep state, even at the lowest packet rate of \SI{200}{packets/\s}. With SR1010, a stable connection could be maintained up to \SI{5.75}{\mega bps}, while with SR1020, the stable connection extended up to \SI{7.24}{\mega bps}. At a data rate of \SI{1.29}{\mega bps}, which is close to the theoretical payload data rate limit of \gls{BLE} 5 \cite{ble5spec}, SR1000 consumes as little as \SI{0.45}{\mW} or \SI{0.35}{\nano\joule /bit} while transmitting and \SI{1.73}{\mW} or \SI{1.34}{\nano\joule /bit} during reception.


\subsection{Communication Range}\label{sub:range}
 

\begin{figure}[!b]
  \vspace*{-1em}
  \center
  \includegraphics[width=0.35\textwidth]{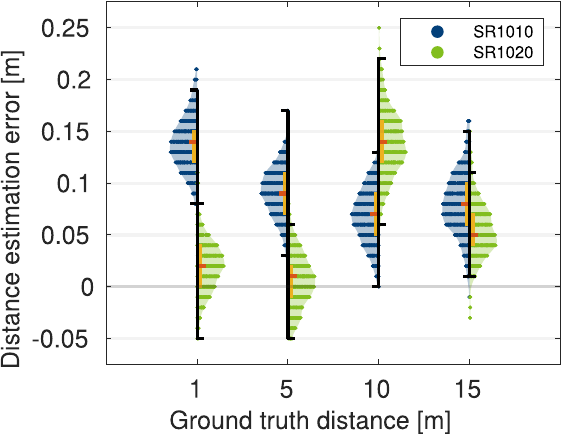}
  \caption{Ranging accuracy for different ground truth distances and the two evaluated \gls{UWB} \glspl{IC}.}
  \label{fig:ranging_violin}
\end{figure}
To assess the maximum usable range, two \glspl{EVK} are placed facing each other in an open field at a height of \SI{1.5}{\m}. One device has been configured to operate as a transmitter, sending a one-byte payload with \gls{2bit-PPM} at a rate of \SI{1000}{packets/\s}. The transmitter counts the number of packets that are successfully acknowledged by the receiving node and calculates the \gls{PDR} over \SI{10}{k} transmissions based on \cref{eq:pdr}. After each measurement, the distance between the two nodes has been increased by \SI{2}{\m}.

\begin{equation}\label{eq:pdr}
    \mathit{PDR} = \frac{
        \text{\# ACKNOWLEDGED PACKETS}   
    }{
        \text{\# SENT PACKETS}
    }
\end{equation}

In \cref{fig:power_res}c), the measured \gls{PDR} for both devices is shown at the two output power settings employed during the power consumption measurements. The SR1010 \gls{PDR} exhibits a significant drop in \gls{PDR} at \SI{30}{\m} and \SI{60}{\m}, followed by a subsequent recovery. This pattern has been observed consistently at three separate measurements. It is most likely caused by multipath reflections, as other measurements with the nodes at the height of \SI{1}{\m} and \SI{2.5}{\m} yielded different results.

\subsection{Time-of-Flight Ranging}\label{sub:ranging}
To analyze the ranging capabilities, the \gls{GUI}'s ranging example is used in the high output power and high precision configuration. The nodes are set up similarly to section \ref{sub:range} but at a distance of \si{1}, \si{5}, \si{10}, and \SI{15}{\m} apart from each other. The precise distance between the nodes has been determined using a \textsc{Leica Disto X310}, which served as the ground truth distance. At each distance, \SI{1}{k} samples have been collected. Prior to the measurement, the transceivers are calibrated according to the application note by \textsc{Spark}.

\cref{fig:ranging_violin} presents the distribution of the measured distance estimation error. The mean error is indicated in orange, while the \SI{50}{\percent} error window is highlighted in yellow. Using the SR1010 transceiver, ranging is achievable up to a distance of \SI{52}{\m}, with SR1020 up to \SI{32}{\m}. The high output power setting in the GUI corresponds to ~\SI{-5.4}{\decibel FS}, which aligns well with the range data measured in \cref{sub:range}.
\FloatBarrier
\section{Conclusion}\label{sec:conclusion}
This paper presented an analysis of the \textsc{Spark Microsystems SR1000} family in the context of low-power, spatial-aware \gls{IoT} applications. This includes a comparison to the most popular \gls{UWB} \glspl{IC} and \gls{BLE}. Experimental evaluations have shown the capabilities of the SR1000 to achieve a high throughput of up to \qty{7.24}{\mega\bit/\second} with a very high bit efficiency of \qty{0.29}{\nano\joule/\bit} when transmitting and \qty{1.39}{\nano\joule/\bit} while receiving. In particular, it proves to be a real alternative to \gls{BLE}: At a data rate of \qty{2}{\mega\bit/\second}, the bit energy efficiency is about \(5\times\) higher than \gls{BLE}, especially for the transmitter at \qty{0.6}{\nano\joule/\bit} compared to \gls{BLE}'s \qty{5.6}{\nano\joule/\bit}. With a ranging accuracy of \(\pm\SI{25}{\centi\meter}\), SR1000 is up to \(2\times\) better than \gls{BLE}, but it falls behind state-of-the-art \gls{UWB} \glspl{IC} which can reach an accuracy of \(\pm\SI{5}{\centi\meter}\). The results show the transceiver's potential to perform well in a wide range of applications due to its low power demand over a wide range of data rates and the ability to perform accurate ranging measurements. 

\section*{Acknowledgments}
The authors would like to thank armasuisse Science \& Technology for funding this research.

\bibliography{./main}

\end{document}